\documentclass[openany,11pt]{article}

\usepackage{pxps}
\usepackage{threeparttable}
\usepackage{float}
\usepackage{graphicx}
\usepackage{subfig}
\usepackage{indentfirst}
\usepackage[affil-it]{authblk}

\def\OMIT#1{}

\def\ifmath#1{\relax\ifmmode#1\else$#1$\fi}

\newcommand{\gsim}{\gtrsim}

\title{{\huge Neutron-Antineutron Oscillations}  \\{\LARGE A Snowmass 2013 White Paper}}
\date{}

\author[16]{K. Babu}
\author[19]{S. Banerjee}
\author[3]{D. V. Baxter}
\author[4,25]{Z. Berezhiani}
\author[26]{M. Bergevin}
\author[19]{S. Bhattacharya}
\author[2]{S. Brice}
\author[15]{T. W. Burgess}
\author[31]{L. Castellanos}
\author[19]{S. Chattopadhyay}
\author[27]{M-C. Chen}
\author[31]{C. E. Coppola}
\author[33]{R. Cowsik}
\author[15]{J. A. Crabtree}
\author[32]{P. Das}
\author[13,22]{E. B. Dees}
\author[5,7,14,24]{A. Dolgov}
\author[12]{G. Dvali}
\author[15]{P. Ferguson}
\author[31]{M. Frost}
\author[31]{T. Gabriel}
\author[18]{A. Gal}
\author[15]{F. Gallmeier}
\author[1]{K. Ganezer}
\author[6]{E. Golubeva}
\author[15]{V. B. Graves}
\author[31]{G. Greene}
\author[31]{T. Handler}
\author[1]{B. Hartfiel}
\author[13]{A. Hawari}
\author[31]{L. Heilbronn}
\author[3]{C. Johnson}
\author[31]{Y. Kamyshkov}
\author[7]{B. Kerbikov}
\author[10]{M. Kitaguchi}
\author[23]{B. Z. Kopeliovich}
\author[6]{V. B. Kopeliovich}
\author[28]{W. Korsch}
\author[6]{V. Kuzmin}
\author[3]{C-Y. Liu}
\author[9]{P. McGaughey}
\author[9]{M. Mocko}
\author[29]{R. Mohapatra}
\author[2]{N. Mokhov}
\author[9]{G. Muhrer}
\author[11]{P. Mumm}
\author[7]{L. Okun}
\author[13,22]{R. W. Pattie Jr.}
\author[13,22,*]{D. G. Phillips II}
\author[2]{C. Quigg}
\author[2]{E. Ramberg}
\author[32]{A. Ray}
\author[8]{A. Roy}
\author[31]{A. Ruggles}
\author[17]{U. Sarkar}
\author[9]{A. Saunders}
\author[20]{A. P. Serebrov}
\author[10]{H. M. Shimizu}
\author[21]{R. Shrock}
\author[32]{A. K. Sikdar}
\author[9]{S. Sjue}
\author[3,*]{W. M. Snow}
\author[2]{A. Soha}
\author[31]{S. Spanier}
\author[2]{S. Striganov}
\author[31]{L. Townsend}
\author[2]{R. Tschirhart}
\author[30]{A. Vainshtein}
\author[3]{R. Van Kooten}
\author[9]{Z. Wang}
\author[13]{B. Wehring}
\author[13,22]{A. R. Young}
\affil[1]{California State University at Dominguez Hills, Carson, CA 90747, USA}
\affil[2]{Fermi National Accelerator Laboratory, Batavia, IL 60510, USA}
\affil[3]{Indiana University, Bloomington, IN 47405, USA}
\affil[4]{INFN, Laboratori Nazionali Gran Sasso, 67100 Assergi, L$'$Aquila, Italy}
\affil[5]{INFN, Sezione di Ferrara, Via Saragat 1, 44122 Ferrara, Italy}
\affil[6]{Institute for Nuclear Research, Russian Academy of Sciences, 117312 Moscow, Russia}
\affil[7]{Institute for Theoretical and Experimental Physics, 113259 Moscow, Russia}
\affil[8]{Inter University Accelerator Centre, New Delhi 110067, India}
\affil[9]{Los Alamos National Laboratory, Los Alamos, NM 87545, USA}
\affil[10]{Nagoya University, Nagoya, Aichi 464-8602, Japan}
\affil[11]{National Institute of Standards and Technology, Gaithersburg, MD 20899, USA}
\affil[12]{New York University, New York, NY 10012, USA}
\affil[13]{North Carolina State University, Raleigh, NC 27695, USA}
\affil[14]{Novosibirsk State University, 630090 Novosibirsk, Russia}
\affil[15]{Oak Ridge National Laboratory, Oak Ridge, TN 37831, USA}
\affil[16]{Oklahoma State University, Stillwater, OK 74074, USA}
\affil[17]{Physical Research Laboratory, Ahmedabad 380009, India}
\affil[18]{Racah Institute of Physics, The Hebrew University, 91904 Jerusalem, Israel}
\affil[19]{Saha Institute of Nuclear Physics, Kolkata 700064, India}
\affil[20]{St. Petersburg Nuclear Physics Institute, Gatchina, 188300 St. Petersburg, Russia}
\affil[21]{State University of New York at Stony Brook, Stony Brook, NY 11790, USA}
\affil[22]{Triangle Universities Nuclear Laboratory, Durham, NC 27710, USA}
\affil[23]{Universidad T\'{e}cnica Federico Santa Mar\'{i}a, Valpara\'{i}so, Chile}
\affil[24]{Universit\`{a} degli Studi di Ferrara, Via Saragat 1, 44122 Ferrara, Italy}
\affil[25]{Universit\`{a} dell$'$Aquila, Via Vetoio, 67100 Coppito, L$'$Aquila, Italy}
\affil[26]{University of California at Davis, Davis, CA 95616, USA}
\affil[27]{University of California at Irvine, Irvine, CA 92697, USA}
\affil[28]{University of Kentucky, Lexington, KY 40506, USA}
\affil[29]{University of Maryland, College Park, MD 20742, USA}
\affil[30]{University of Minnesota, Minneapolis, MN 55455, USA} 
\affil[31]{University of Tennessee, Knoxville, TN 37996, USA}
\affil[32]{Variable Energy Cyclotron Centre, Kolkata 700064, India}
\affil[33]{Washington University, St. Louis, MO 63130, USA}

\begin{document}
\maketitle

\vspace{-0.5 in}

\let\oldthefootnote\thefootnote
\renewcommand{\thefootnote}{\fnsymbol{footnote}}
\footnotetext[1]{To whom correspondence should be addressed. W. M. Snow: \url{wsnow@indiana.edu}, D. G. Phillips II \url{dgphilli@ncsu.edu}}
\let\thefootnote\oldthefootnote

\vspace{+0.1 in}

\begin{abstract}
\phantom \noindent This paper summarizes discussions of the theoretical developments and the studies performed by the NNbarX collaboration for the 2013 Snowmass Community Summer Study. 
\end{abstract}
\clearpage
\section{Executive Summary}
\label{nnbar:sec:execsummary}

The discovery of neutrons turning into antineutrons would have a significant impact on the world of particle physics
and cosmology by demonstrating that baryon number ($\mathcal{B}$) is not conserved and showing that
all matter containing neutrons is unstable. It would imply that the matter in our universe
can evolve from the initial $\mathcal{B}$=0 void predicted by inflation and thereby answer
the very fundamental cosmological question of the origin of the observed matter-antimatter
asymmetry of the universe. By showing $\mathcal{B}$ is violated by 2 units, its discovery would
strongly suggest that the physics of quark-lepton unification and neutrino mass generation is near
the TeV scale with far reaching implications for Large Hadron Collider (LHC) searches. If seen at rates observable in
a foreseeable next-generation experiment, its effects must be taken into account for any
quantitative understanding of the baryon asymmetry of the universe. The experimental signature
of antineutron annihilation in a free neutron beam is spectacular enough that an essentially
``background free'' search is possible, while any positive observation can be turned off by
a very small change in the experiment's ambient magnetic field. An optimized experimental search
for oscillations using free neutrons from a 1 MW spallation target at Fermi National Accelerator Laboratory's (FNAL) Project X~\cite{Kronfeld:2013ak} can
improve existing limits on the free oscillation probability by 4 orders of magnitude by fully
exploiting new slow neutron source and optics technology developed for materials research in
an experiment delivering a slow neutron beam through a magnetically-shielded vacuum to a thin
annihilation target. A null result at this level would represent the most stringent limit on
matter instability above 10$^{35}$ $\rm{yrs}$~\cite{Dover:1983cd,Friedman:2008ef}. Combined
with data from the LHC and other searches for rare processes, a null result could also rule out
a scenario for baryogenesis below the electroweak phase transition. 

\section{Physics Motivation for $n - \bar n$ Searches}
\label{nnbar:sec:physics}

Historically, the idea that neutron and antineutron can be states belonging to the same particle was first conjectured in 1937~\cite{Majorana:1937em}.  Although particle physics since that time has witnessed the success of Quantum Chromodynamics and has evolved to accept $\mathcal{B}$ as a good symmetry to understand observed nuclear phenomena, a tiny Majorana component to the neutron mass that violates $\mathcal{B}$ still remains an intriguing possibility with far reaching implications.  The early history of other physics ideas related to $n$-$\bar{n}$ oscillations is briefly discussed elsewhere~\cite{Okun:2013lo}.

There are many compelling reasons to think that fundamental particle interactions violate
$\mathcal{B}$. Arguably, the most powerful reason is that generating the observed matter-antimatter asymmetry in the universe requires that $\mathcal{B}$ must be violated~\cite{Sakharov:1967as}.  Cosmological inflation, which is strongly supported by astronomical data, coupled with the fact that the universe has an excess of matter over antimatter, implies that baryon number ($\mathcal{B}$) must be violated~\cite{Dolgov:1992ad,Dolgov:1998ad}.  Other reasons include grand unified theories
(GUTs)~\cite{Georgi:1974hg,Raby:2008sr} and non-perturbative effects in the Standard Model
itself, which lead to $\mathcal{B}$-violation ~\cite{Hooft:1976gh,Hooft:1978gh,Kuzmin:1985vk}.
The $\mathcal{B}$-violating effects in all these models appear with a very weak strength
so that stability of atoms such as hydrogen and helium are not significantly affected
on the time scale of the age of the universe. 

Once we accept the possibility that $\mathcal{B}$ is not a good symmetry of nature, there
are many questions that must be explored to decide the nature of physics associated with
$\mathcal{B}$-violation:  is (a non-anomalous extension of) baryon number, $\mathcal{B}$,
a global or local symmetry?  Does $\mathcal{B}$ occur as a symmetry by itself or does it appear in combination with
lepton number, $\mathcal{L}$, i.e. $\mathcal{B}$ - $\mathcal{L}$, as the Standard Model
(SM) would suggest? What is the scale of $\mathcal{B}$-violation and the nature of the
associated physics that is responsible for it? For example, is this physics characterized
by a mass scale not too far above the TeV scale, so that it can be probed in experiments
already searching for new physics in colliders as well as in low-energy rare processes? Are
the details of the physics responsible for $\mathcal{B}$-violation such that they can
explain the origin of matter?

Proton decay searches probe $\mathcal{B}$-violation due to physics at a grand
unified scale of $\sim 10^{15}-10^{16}$ GeV.  In contrast, the $\mathcal{B}$-violating
process of $n$-$\bar{n}$ oscillation, where a free neutron spontaneously
transmutes itself into an antineutron, has very different properties and
probes quite different physics.  The oscillation process violates $\mathcal{B}$ by two units and is caused by a dimension nine operator if only standard model fields are involved. In this case, it probes mass scales $\sim$ 100 TeV if all couplings are assumed to be of order $\sim$ 1.  However, examples of specific theories exist where new beyond-standard-model (BSM) particles with smaller couplings to quarks are predicted by other considerations e.g. baryogenesis,  so that new particles with masses near a TeV scale mediating neutron oscillation can exist. These theories can be probed at the LHC, providing complementary information on the models for neutron oscillation.  It may also be deeply
connected to the possibility that neutrinos may be Majorana fermions, a natural
expectation.  A key question for experiments is whether there are theories that
predict $n$-$\bar{n}$ oscillations at a level that can be probed in
currently available facilities such as reactors or in contemplated ones such as
Project X, with intense neutron fluxes.  Equally important are the constraints that would be imposed on models for BSM physics should no evidence for oscillations be seen within the projected reach of a next generation of free-neutron searches (from 10$^{9}$ to 10$^{10}$ s).

Motivated by theoretical possibility of baryon number violation 
by two units several experimental searches~\cite{Beringer:2012jb}
have been performed with free neutrons in vacuum~\cite{Baldo:1994bc} and with neutrons bound inside nuclei~\cite{Beringer:2012jb} that set the limits for the free neutron-
antineutron oscillation time at the level $\gsim$ 10$^{8}$ s. These experiments
and corresponding limits are discussed further in our paper.

%
\subsection{Some Background Concerning Baryon Number Violation} 
\label{nnbar:subsec:theorybkgd}

Early on, it was observed that in a model with a left-right symmetric
electroweak group, $G_{LR} = {\rm SU}(2)_L \otimes {\rm SU}(2)_R \otimes {\rm
U}(1)_{B-L}$, baryon and lepton numbers in the combination $\mathcal{B}$ - $\mathcal{L}$ can be
gauged in an anomaly-free manner~\cite{Pati:1974jp,Mohapatra:1975rm,Mohapatra:1975jp,Senjanovic:1975gs}.  The resultant U(1)$_{B-L}$ can be combined
with color SU(3) in an SU(4) gauge group \cite{Pati:1974jp}, giving rise to the group
$G_{422} = {\rm SU}(4) \otimes {\rm SU}(2)_L \otimes {\rm SU}(2)_R$.  A higher degree of unification involved models that embed either
the Standard Model gauge group $G_{SM} = {\rm SU}(3)_c \otimes {\rm SU}(2)_L
\otimes {\rm U}(1)_Y$ or $G_{422}$ in a simple group such as SU(5) or SO(10)
~\cite{Georgi:1974hg,Raby:2008sr}.  The motivations for grand unification theories
are well-known and include the unification of gauge interactions and their
couplings, the related explanation of the quantization of weak hypercharge and
electric charge, and the unification of quarks and leptons. While the gauge
couplings do not unify in the Standard Model, they do unify in a minimal
supersymmetric extension of the Standard Model.  Although supersymmetric
particles have not been discovered in the 7 TeV and 8 TeV data at the Large
Hadron Collider, they may still be observed at higher energy.  Supersymmetric
grand unified theories thus provide an appealing possible ultraviolet
completion of the Standard Model. The unification of quarks and leptons in
grand unified theories (GUTs) generically leads to the decay of the protons and neutrons in nuclei with violation of baryon number.  These
decays typically obey the selection rule $\Delta \mathcal{B} = -1$ and $\Delta \mathcal{L} = -1$.
However, the general possibility of a different kind of baryon-number violating
process, namely the $|\Delta \mathcal{B}|=2$ process of $n - \bar n$ oscillations, was
suggested~\cite{Kuzmin:1970vk} even before the advent of GUTs.  This was further
discussed and studied after the development of GUTs in~\cite{Glashow:1979sg,Mohapatra:1980rm} and
in a number of subsequent models~\cite{Kuo:1980tk,Chang:1980lc,Mohapatra:1980rn,Cowsik:1981rc,Rao:1982sr,
Misra:1983sm,Rao:1984sr,Huber:2001sh,Babu:2001kb,Nussinov:2002sn,Babu:2006kb,Dutta:2006bd,Babu:2009kb,
Mohapatra:2009rm,Gu:2011pg,Babu:2013kb,Arnold:2013ja,Winslow:2010pw}.
Recently, a number of models have been constructed that predict $n - \bar n$ oscillations
at levels within reach of possible new experimental searches~\cite{Babu:2001kb,Nussinov:2002sn,Dutta:2006bd,Babu:2013kb,Winslow:2010pw}.


\subsection{$n- \bar n$ Oscillations in Vacuum}
\label{nnbar:subsec:formalism}

Since the neutron and antineutron have opposite magnetic moments, one must account for the magnetic splittings that may
be present between $n$ and $\bar{n}$ states in an oscillation experiment.  This motivates the following review
of the formalism for the two level ($n$,$\bar{n}$) system and $n - \bar n$ oscillations in an external magnetic
field~\cite{Mohapatra:1980rn,Cowsik:1981rc}.  

The $n$ and $\bar n$ interact with the external $\vec B$ field via their
magnetic dipole moments, ${\vec \mu}_{n,\bar n}$, where 
$\mu_n = -\mu_{\bar n} = -1.9 \mu_N$ and $\mu_N = e/(2m_N) = 
3.15 \times 10^{-14}$ MeV/Tesla.  Hence, the effective Hamiltonian matrix for the two-level $n - \bar n$ system takes the form 

\begin{equation}
\cal{M}_{B}=\left(\begin{array}{cc}
m_n - {\vec \mu}_n \cdot {\vec B} - i\lambda/2 & \delta m \\
\delta m         & m_n + {\vec \mu}_n \cdot {\vec B} -  i\lambda/2
\end{array}\right),
\numberwithin{equation}{section}
\end{equation}
where $m_{n}$ is the mass of the neutron, $\delta m$ is the $\mathcal{B}$-violating potential coupling the $n$
and $\bar{n}$ states, and 1/$\lambda$ = $\tau_{n}$ = (880.0$~\pm$ 0.9) $\rm{s}$~\cite{Beringer:2012jb} is the mean neutron lifetime.  

The transition probability for a neutron traveling in vacuum for a time $t$ since its last interaction with matter is given by
$P(n(t)=\bar n) = \sin^2(2\theta) \, \sin^2 [(\Delta E)t/2] \, e^{-\lambda t}$, 
where $\Delta E \simeq 2 |\vec{\mu}_n\cdot\vec{B}|$ and $\tan(2\theta) = - \delta m/(\vec{\mu}_n\cdot\vec{B})$.
In a free propagation experiment, the quasi-free condition must hold, such that $|{\vec \mu}_n \cdot {\vec B}|t\ll1$.
In this limit and also assuming that $t\ll \tau_{n}$, $P(n(t)=\bar n) \simeq [(\delta m) \, t]^2 = (t/\tau_{n- \bar n})^2$. 

The number of $\bar n$'s produced by the $n - \bar n$ oscillations is given by $N_{\bar n}=P(n(t)=\bar n)N_n$, where $N_n = \phi T_{run}$, with
$\phi$ the integrated neutron flux and $T_{run}$ the running time.  When the quasi-free condition holds, the sensitivity of the
experiment depends in part on the product $t^2 \phi$, so, with adequate
magnetic shielding, one wants to maximize $t$, subject to the condition that 
$|{\vec \mu}_n \cdot {\vec B}|t\ll 1$.


\subsection{$n - \bar n$ Oscillations in Nuclei} 
\label{nnbar:subsec:matter}

To put the proposed free propagation $n$-$\bar{n}$ oscillation experiment in
perspective, it is appropriate to review limits that have been achieved in the
search for $n - \bar n$ oscillations in nuclei, using large nucleon decay
detectors~\cite{Beringer:2012jb}.  

Most of the neutrons available for experiments are contained 
inside nuclei.  However, their transformation to antineutrons 
is heavily suppressed by the nuclear potential difference for particle 
and antiparticle components of the neutron wave function.  Large 
magnitudes of suppression can be qualitatively explained by a simple
argument: neutrons inside nuclei are quasi-free for the time 
$\Delta t\sim 1/\Delta E$, where $\Delta E$ is of the order of the neutron 
binding energy inside the nuclei, e.g. $\sim$ 30 MeV. During the time $\Delta t$, the 
quasi-free neutron acquires a probability of transformation to an antineutron
proportional to $\Delta t^{2}$.  At the end of the period $\Delta t$, the wavefunction 
is reset due to different interactions within nuclear potentials of different 
components of the wavefunction. This quasi-free condition occurs in nuclei 
1/$\Delta t$ times per second. Thus, the total width for an intranuclear $n$-$\bar{n}$ 
transformation is equal to $\Delta t$/$\tau_{n- \bar n}^{2}$ and the intranuclear lifetime $\tau_{m}$ 
in respect to $n$-$\bar{n}$ transformation is $\tau_{m} = R\cdot \tau_{n- \bar n}^{2}$, where R is the ``nuclear suppression factor" approximately equal to 1/$\Delta t$.
Several more elaborate nuclear model calculations give results within the 
same order of magnitude~\cite{Dover:1983cd,Friedman:2008ef,Vainshtein:2013av}, and slightly different for different nuclei. 
The nuclear suppression factor is somewhat smaller for deuterium due to the small 
binding energy. 

The best limit for an intranuclear $n$-$\bar{n}$ search was published in 2002
by the Soudan II collaboration for $^{56}$Fe nuclei as $\tau_m > 0.72 \times 10^{32} \ {\rm yr} \ (90 \% \  {\rm CL})$~\cite{Chung:2002jc}.  Using a suppression factor of $R \simeq 1.4\times10^{-23}$~s$^{-1}$~\cite{Dover:1983cd}, this 
corresponds to a free $n$-$\bar{n}$ oscillation time of $\tau_{n- \bar n} > 1.3 \times 10^8$~s.
Preliminary results of the SNO Collaboration on deuterium and $^{16}$O using a
fraction of available experimental statistics~\cite{Bergevin:2010mb} together with the most recent 
suppression factor for deuterium~\cite{Kopeliovich:2011vk} gives a new limit 
for a free $n$-$\bar{n}$ oscillation time of $\tau_{n- \bar n} > 1.8 \times 10^8$~s.
The preliminary result from the Super-Kamiokande collaboration for the $^{16}$O 
nuclei lifetime (for $n$-$\bar{n}$) is $\tau_m > 1.9  \times 10^{32} \ {\rm yr} \ (90 \% \ {\rm CL})$~\cite{Abe:2011ka}, which is translated~\cite{Dover:1983cd}
to a free neutron oscillation time of $\tau_{n- \bar n} > 2.44\times 10^8$ s.  However,
with the more advanced nuclear suppression factor in~\cite{Friedman:2008ef} the same Super-Kamiokande result should 
correspond to $\tau_{n- \bar n} > 3.5 \times 10^8$ s.

Common for the limits in all of the three intranuclear experiments mentioned 
above is the presence of irreducible backgrounds associated with detection of 
atmospheric neutrinos in deep underground detectors.  The limit from the Super-Kamiokande 
collaboration is based on 24 candidate events with an expected calculated 
background of 24.1 events.  If this situation can not be improved with large 
detectors employing new technologies, like for example in a large liquid Argon (LAr) detector where the atmospheric 
neutrino background is expected to be significantly suppressed by additional 
high-resolution spatial and ionization information, then the background will 
remain a factor limiting further progress for $n$-$\bar{n}$ search in intranuclear 
$n$-$\bar{n}$ transformations.  The advantages of LAr detectors for suppressing 
atmospheric neutrino backgrounds in the detection of $n$-$\bar{n}$ events still 
remains to be experimentally demonstrated.


\subsection{Operator Analysis and Estimate of Matrix Elements}
\label{nnbar:subsec:analysis}

At the quark level, the $n \to \bar n$ transition is 
$(u d d) \to (u^c d^c d^c)$.  This is
mediated by six-quark operators ${\cal O}_i$, so the transition amplitude is
characterized by an effective mass scale $M_X$ and is expressed as
\begin{equation}
\delta m = \langle \bar n | H_{eff} | n \rangle = \frac{1}{M_X^5}
\sum_i c_i \langle \bar n |{\cal O}_i  | n \rangle .
\numberwithin{equation}{section}
\end{equation}
Hence,
$\delta m \sim \kappa \Lambda_{QCD}^{6}/M_X^{5}$, 
where $\kappa$ is a generic $\kappa_i$ and $\Lambda_{QCD} \simeq 200$ MeV
arises from the matrix element $\langle \bar n | {\cal O}_i | n \rangle$.  For
$M_X$ of order 10$^5$ GeV, one has $\tau_{n- \bar n} \simeq 10^9 \
{\rm s}$.

The operators ${\cal O}_i$ must be color singlets and, for $M_X$
larger than the electroweak symmetry breaking scale, also ${\rm SU}(2)_L \times
{\rm U}(1)_Y$-singlets.  An analysis of these (operators) was carried out in~\cite{Rao:1982sr}
and the $\langle \bar n | {\cal O}_i | n \rangle$ matrix elements were 
calculated in the MIT bag model.  Further results were obtained varying MIT bag
model parameters in~\cite{Rao:1984sr}.  These calculations involve integrals over 
sixth-power polynomials of spherical Bessel functions from the quark 
wavefunctions in the bag model.  From the arguments above,
it was found that 
\begin{equation}
|\langle \bar n | {\cal O}_i | n \rangle | \sim O(10^{-4})
\ {\rm GeV}^6 \simeq (200 \ {\rm MeV})^6 \simeq \Lambda_{QCD}^6
\numberwithin{equation}{section}
\end{equation}
An exploratory effort has recently begun to calculate these matrix elements
using lattice gauge theory methods \cite{Buchoff:2012mb}.  Given that the mass scales probed
by these measurements go well beyond the TeV scale, the fundamental impact of a result (whether
or not oscillations are observed) and the availability of a variety of models predicting $n$-$\bar{n}$ at current sensitivity
levels ($\tau_{n- \bar n}\sim 10^{8}$ s), there is strong motivation to pursue a higher-sensitivity
$n - \bar n$ oscillation search experiment that can achieve a lower bound of $\tau_{n- \bar n} \sim
10^9 - 10^{10}$ s.  

\section{NNbarX: An Experimental Search for $n- \bar n$ Oscillations at FNAL}
\label{nnbar:sec:neutronpx}

Project X presents an opportunity to probe $n$-$\bar{n}$ transformation
with free neutrons with an unprecedented improvement in sensitivity~\cite{Kronfeld:2013ak}.  Improvements would
be achieved by creating a unique facility, combining a high intensity
cold neutron source {\it dedicated} to particle physics experiments with
advanced neutron optics technology and detectors which build on
the demonstrated capability to detect antineutron
annihilation events with zero background.  Existing slow neutron sources at research reactors
and spallation sources possess neither the required space nor the degree of access
to the cold source needed to take full advantage of advanced neutron
optics technology which enables a greatly improved free $n$-$\bar{n}$ transformation
search experiment. Therefore, a dedicated source devoted exclusively
to fundamental neutron physics, such as would be available at Project
X, represents an exciting tool to explore not only $n$-$\bar{n}$ oscillations, 
but also other Intensity Frontier questions accessible through slow neutrons.

\subsection{Previous Experimental Searches for $n- \bar n$ Oscillations}
\label{nnbar:subsec:previous}

As mentioned in Sec.~\ref{nnbar:subsec:matter}, the current best limit on $n$-$\bar{n}$ oscillations comes from the
Super-Kamiokande experiment, which determined an upper-bound on the
free neutron oscillation time of $\tau_{n-\bar{n}} >$ 3.5$\times10^{8}$ s
from $n$-$\bar{n}$ transformation in $^{16}$O nuclei~\cite{Friedman:2008ef,Abe:2011ka}.
An important point for underground water Cherenkov measurements is that these experiments
are already limited in part by atmospheric
neutrino backgrounds.  Because only modest increments in detector mass
over Super-Kamiokande are feasible and the atmospheric neutrino backgrounds will scale
with the detector mass, dramatic improvements in the current limit
will be unrealizable for such experiments.

Experiments which utilize free neutrons to search for $n$-$\bar{n}$
oscillations have a number of remarkable features.  The basic idea for
these experiments is to prepare a beam of slow (below room temperature) neutrons which propagate
freely from the exit of a neutron guide to a distant annihilation target.  During
the time in which the neutron propagates freely, a $\mathcal{B}$-violating interaction can
produce oscillations from a pure ``$n$" state to one with an admixture of ``$n$" and ``${\bar n}$" amplitudes. Antineutron appearance is sought through
annihilation in a thin target, which generates several secondary
pions seen by a tracking detector situated around the target.  This signature
strongly suppresses backgrounds.  

To observe an $n$-$\bar{n}$ oscillation signal using free neutrons, the ``quasi-free" condition must hold, in which the $n$ and $\bar{n}$
are effectively degenerate.  This creates a requirement for low pressures
(below roughly $10^{-5}$ Pa for Project X) and very small ambient magnetic fields
(between 1 and 10 nT for Project X) in order to prevent splittings between the
neutron and antineutron from damping the oscillations.  This feature also provides a unique and robust test for a ``false positive" result due to backgrounds.    By deliberately lifting the quasi-free condition (introduce a mT ambient magnetic field to the neutron beam drift region), one can effectively eliminate any antineutron annihilation events in the target, permitting a rigorous evaluation of annihilation detector backgrounds.  An improvement in sensitivity over the current free-neutron limit is available through the use of cutting-edge neutron optics, greatly increasing the neutron integrated flux and average transit time to the annihilation target.  

\begin{figure}
    \centering \includegraphics[width=0.92\textwidth]{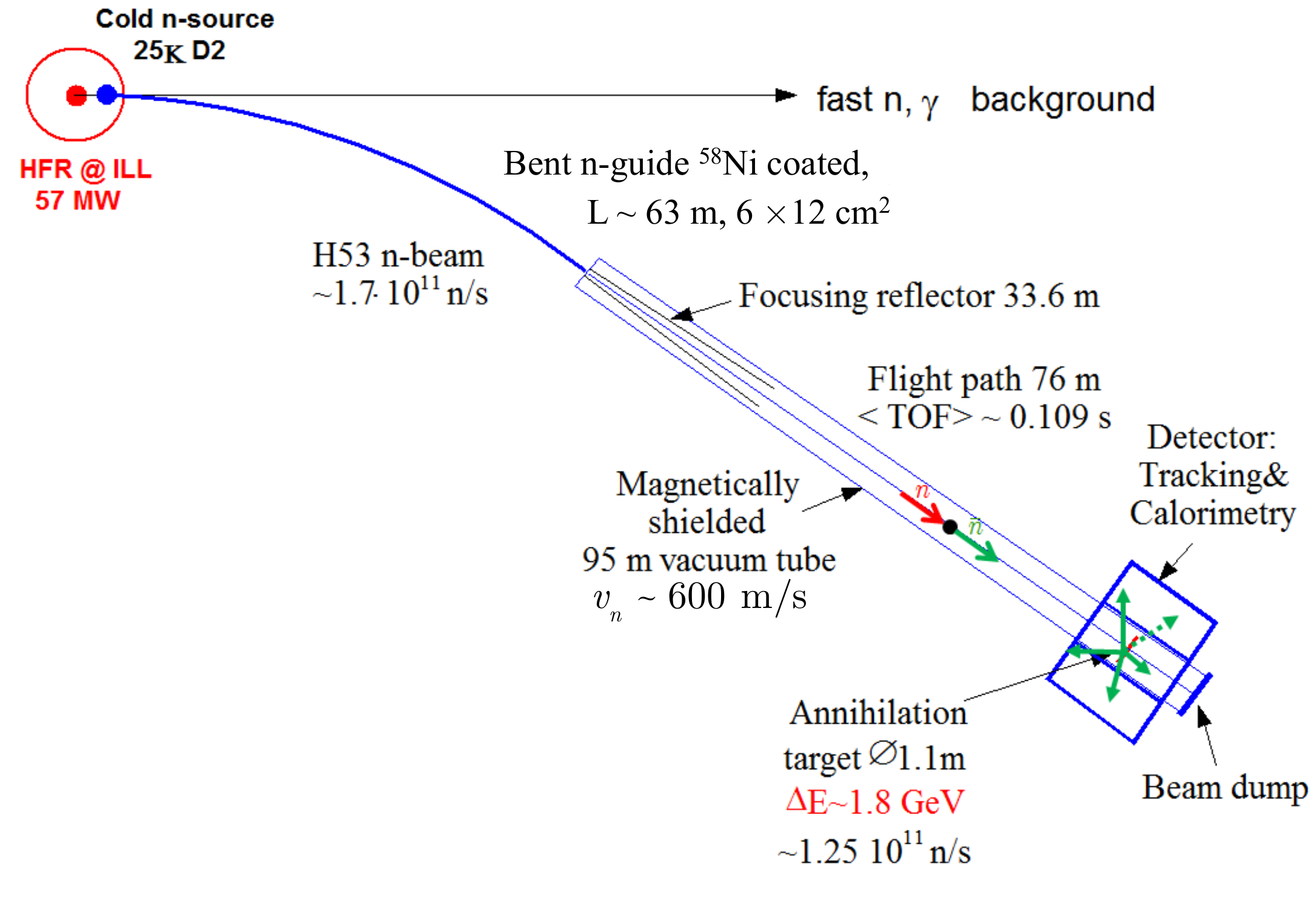} 
    \caption{Configuration of the horizontal $n$-$\bar{n}$ search
    experiment at ILL/Grenoble~\cite{Baldo:1994bc}.}
    \label{ill:fig:logo}
\end{figure}

The current best limit for an experimental search for free $n$-$\bar{n}$
oscillations was performed at the ILL in
Grenoble from 1989 - 1991~\cite{Baldo:1994bc} (see Fig.~\ref{ill:fig:logo}).
This experiment used a cold neutron beam from their 57 MW research reactor
with a neutron current of 1.25$\times$10$^{11} {\it n}/{\rm s}$
incident on the annihilation target and gave a limit of
$\tau_{n-\bar{n}} > 0.86\times10^{8}$ $\rm{s}$~\cite{Baldo:1994bc}.
The neutrons from the source were moderated to a temperature of 25 K, which corresponded to an average velocity of $\sim$ 600 ${\rm m/s}$ and the average neutron observation time was $t_{RMS}\sim$ 0.109 ${\rm s}$~\cite{Bitter:1992tb}. A vacuum of $P\simeq 2\times10^{-4}$ ${\rm Pa}$ maintained in the
neutron flight volume and a magnetic field of $|{\vec B}| < 10$ ${\rm nT}$
satisfied the quasi-free conditions for oscillations to occur.
Antineutron appearance was sought through annihilation with a $\sim$
130 ${\rm \mu m}$ thick carbon film target which generated at least
two tracks (one due to a charged particle) in the tracking detector with
a total energy above 850 MeV in the surrounding calorimeter. In one year of
operation the ILL experiment saw zero candidate events with zero
background~\cite{Baldo:1994bc}.

\subsection{Overview of the NNbarX Experiment}
\label{nnbar:subsec:overview}

A $n$-$\bar{n}$ oscillation search experiment at Project X (NNbarX) is
conceived of as a two-stage experiment. The neutron spallation
target/moderator/reflector system and the experimental apparatus need
to be designed together in order to optimize the sensitivity of the
experiment. The target system and the first-stage experiment can be
built and start operation during the commissioning of the first-stage
of Project X, which is based on a 1 GeV proton beam Linac
operating at 1 mA. The first-stage of NNbarX will be a horizontal
experiment with configuration similar to the ILL experiment~\cite{Baldo:1994bc}, but employing
modernized technologies which include an optimized slow neutron
target/moderator/reflector system and an elliptical supermirror
neutron focusing reflector. Our very conservative baseline goal for a
first-stage experiment is a factor of 30 improvement of the
sensitivity, $N_{n}\cdot t^{2}$, for $n$-$\bar{n}$ oscillations
beyond the limits obtained in the ILL experiment~\cite{Baldo:1994bc}, where $N_{n}$
is the number of free neutrons observed and ${\it t}$ is the neutron
observation time (discussed in Sec.~\ref{nnbar:subsec:formalism}). 
This level of sensitivity would also surpass the $n$-$\bar{n}$
oscillation limits obtained in the Super-Kamiokande, Soudan-II, and SNO intranuclear
searches~\cite{Abe:2011ka,Chung:2002jc,Bergevin:2010mb}.  In fact, although still in progress, our
optimization studies indicate that this horizontal geometry is capable
of improvements of a factor of 300 or more in sensitivity over 3 years of operation at
Project X.  A future, second stage of an NNbarX experiment can achieve higher
sensitivity by exploiting a vertical layout and a moderator/reflector
system which can make use of colder neutrons and ultracold neutrons
(UCN) for the $n$-$\bar{n}$ search.  This experimental arrangement
involves new technologies that will require a dedicated R$\&$D
campaign, but the sensitivity of NNbarX should improve by another
factor of $\sim$ 100 with this configuration, corresponding to limits
for the oscillation time parameter $\tau_{n-\bar{n}} > 10^{10}$
$\rm{s}$.  The increased sensitivity for $n$-$\bar{n}$ oscillations beyond all current experimental limits~\cite{Baldo:1994bc,Abe:2011ka,Chung:2002jc,Bergevin:2010mb} provide a strong motivation to search
for $n$-$\bar{n}$ oscillations as a part of Project X.

The reason there has been no improvement
in the limit on free neutron $n$-$\bar{n}$ oscillations since the ILL
experiment is that no substantial improvement is possible
using existing sources.  Intense beams of very low energy neutrons (meV) are available at
facilities optimized for condensed matter studies focused on neutron
scattering. These sources may be based on high flux reactors such as
the ILL or the High Flux Isotope Reactor (Oak Ridge) or on accelerator
based spallation sources such as SINQ
(Switzerland)~\cite{Blau:2009bb,Fischer:1997wf}, the SNS~\cite{Mason:2006tm},
or the JSNS in Japan~\cite{Maekawa:2010fk}.  Existing neutron
sources are designed and optimized to serve a large number of neutron scattering instruments, each of which requires a beam with a relatively small cross-sectional area.  A fully optimized neutron source
for an $n$-$\bar{n}$ oscillation experiment would require a beam having
a very large cross section and large solid angle. There are no such
beams at existing sources as these attributes would preclude them from providing the resolution
necessary for virtually all instruments suitable for materials
research. The creation of such a beam at an existing facility would
require major modifications to the source/moderator/shielding
configuration that would seriously impact the efficacy for neutron
scattering.  

The initial intensity of the neutron source was determined in the ILL
experiment by the brightness of the liquid deuterium cold neutron source
and the transmission of the curved neutron guide.  Although one expects the
sensitivity to improve as the average velocity of neutrons is reduced, it is
not practical to use very cold neutrons ($<$ 200 ${\rm m/s}$) with a horizontal
layout for the $n$-$\bar{n}$ search due to effects of Earth's gravity,
which will not allow free transport of very slow neutrons over significant
distances in the horizontal direction. Modest improvements in the magnetic
field and vacuum levels reached for the ILL experiment would still
assure satisfaction of the quasi-free condition for the horizontal
experiment planned at Project X, but in our ongoing optimizations we
will investigate limits of $|{\vec B}|\leq 1$ ${\rm nT}$ in the whole free
flight volume and vacuum better than $P\sim 10^{-5}$ ${\rm Pa}$ in
anticipation of the more stringent requirements for a vertical
experiment. The costs of realizing these more stringent goals will be
considered in our optimization of the experimental design.

\begin{figure}[hbtp]
  \centering
    \subfloat[]{\includegraphics[width=4.0in]{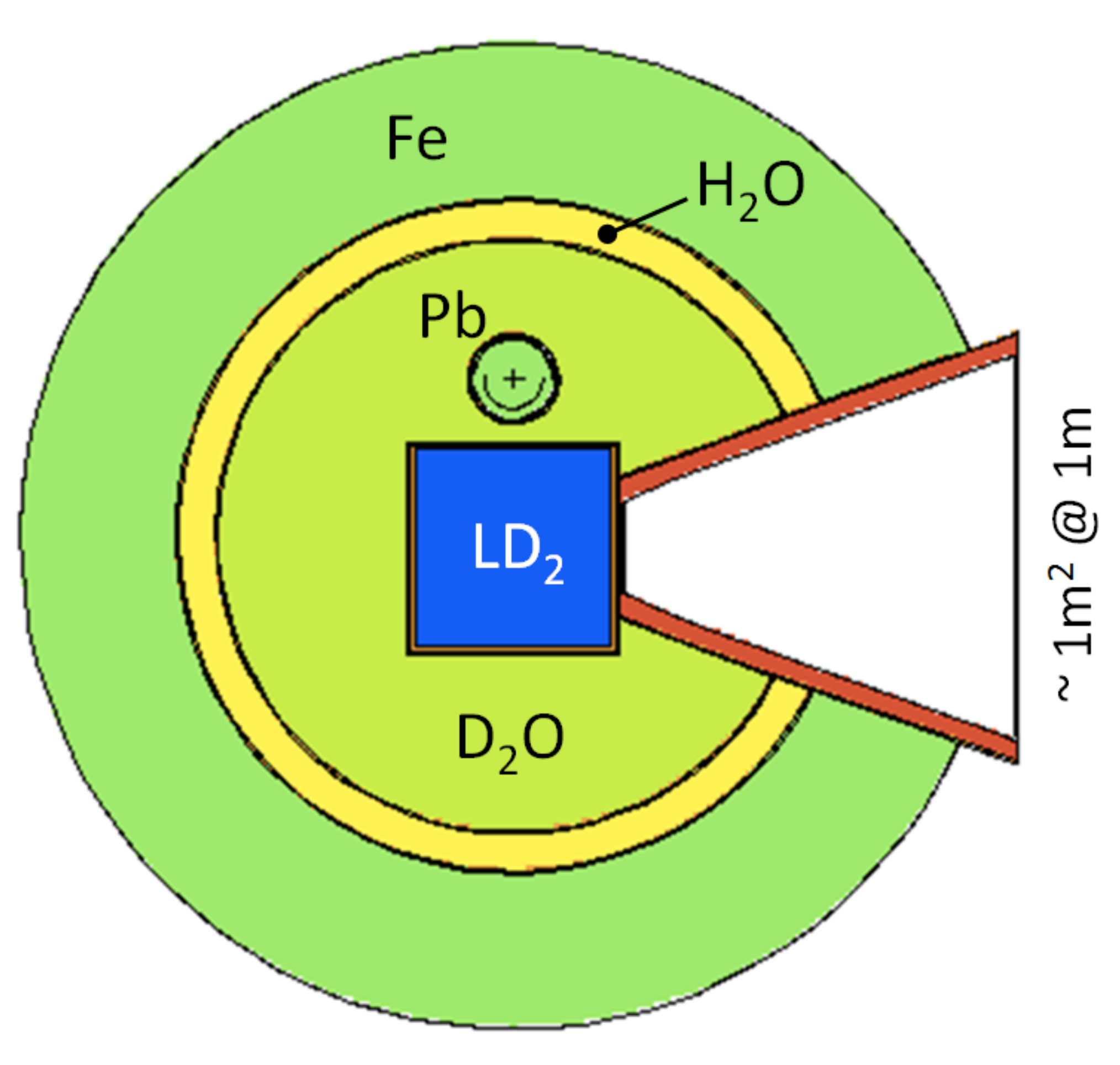}}
    \qquad
    \subfloat[]{\includegraphics[width=5.0in]{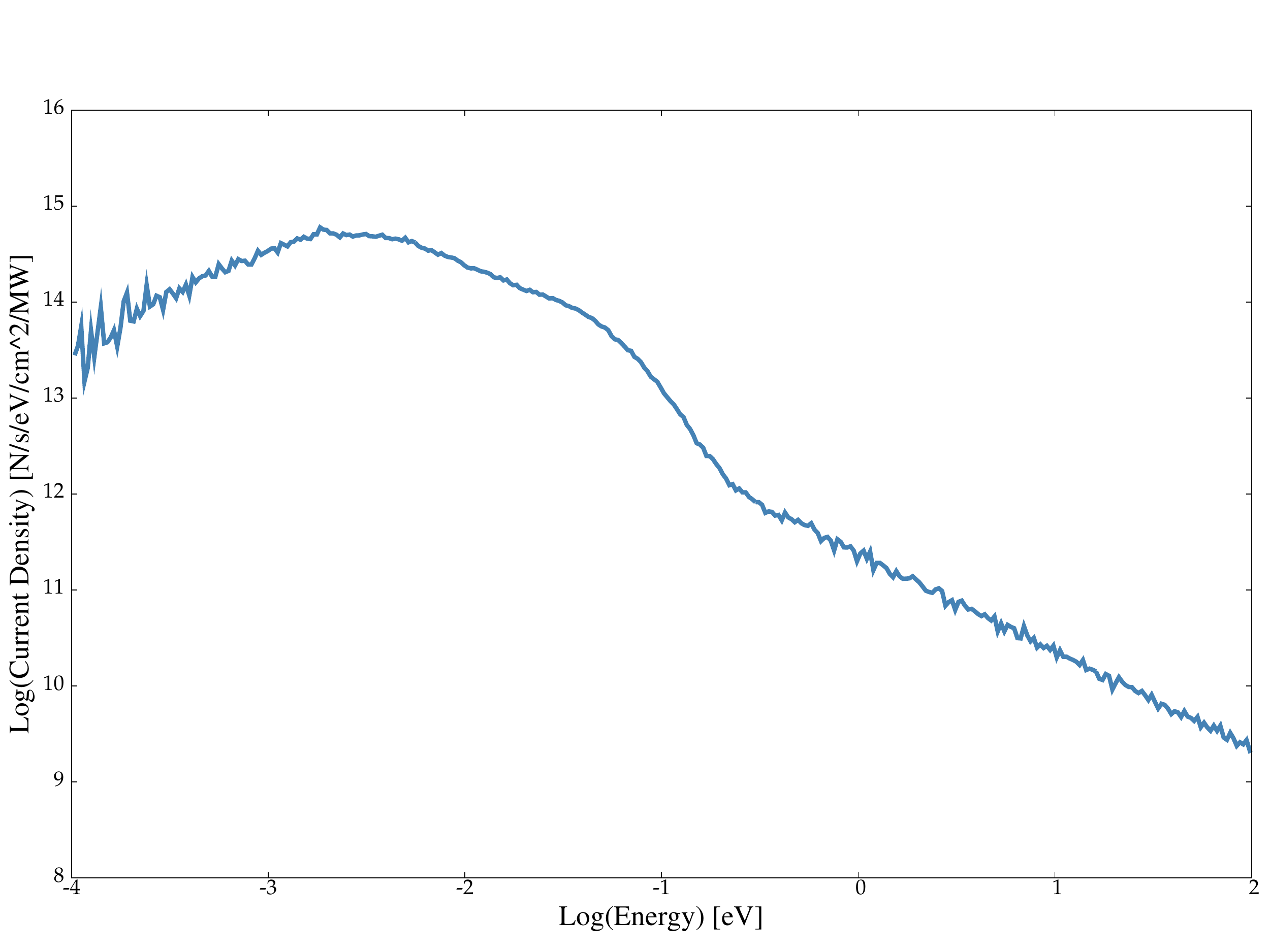}}
    \caption{a) Depiction of the initial NNbarX baseline
cold neutron source geometry.  b) MCNPX simulation of the cold neutron
spectrum entering the neutron optical system.}
    \label{nnbarx:fig:source}
\end{figure}

The Project X spallation target system will include a cooled
spallation target, reflectors and cold source cryogenics, remote
handling, non-conventional utilities, and shielding.  The delivery
point of any high-intensity beam is a target which presents
technically challenging issues for optimized engineering design, in
that optimal neutron performance must be balanced by effective
strategies for heat removal, radiation damage, remote handling of
radioactive target elements, shielding, and other aspects and
components of reliable safe operation.  The NNbarX baseline design incorporates
a spallation target core, which can be cooled by circulating water or
heavy water and will be coupled to a liquid deuterium cryogenic moderator
with optimized size and performance (see Fig. \ref{nnbarx:fig:source}).  

\subsection{Increased Sensitivity of the NNbarX Experiment}
\label{nnbar:subsec:sensitivity}

A higher sensitivity in the NNbarX experiment compared to the previous
ILL experiment~\cite{Baldo:1994bc}, can be achieved by employing
various improvements in neutron optics and
moderation~\cite{Snow:2009ws}.  Conventional moderator designs can be
enhanced to increase the yield of cold neutrons through a number of
neutronics techniques such as a re-entrant moderator
design~\cite{Ageron:1989pa}, use of
reflector/filters~\cite{Mocko:2013mm},  supermirror
reflectors~\cite{Swiss:2013sn}, and high-albedo materials such as
diamond nanoparticle
composites~\cite{Nezvizhevsky:2008vz,Lychagin:2009el,Lychagin:2009em}.
Although potentially of high positive impact for an $n$-$\bar{n}$
experiment, some of these techniques are not necessarily suitable for
multi-purpose spallation sources serving a materials research user
community (where sharply defined neutron pulses in time may be
required, for example).

\begin{figure}[hbtp]
  \centering
    \subfloat[]{\includegraphics[width=5.3in]{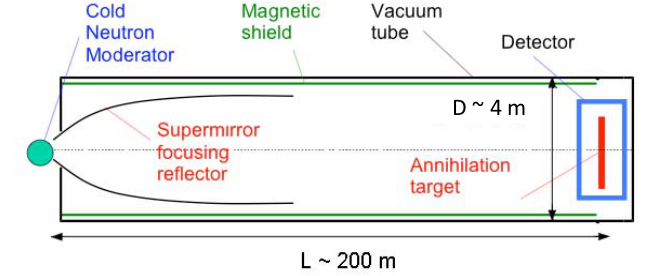}}
    \qquad
    \subfloat[]{\includegraphics[width=5.0in]{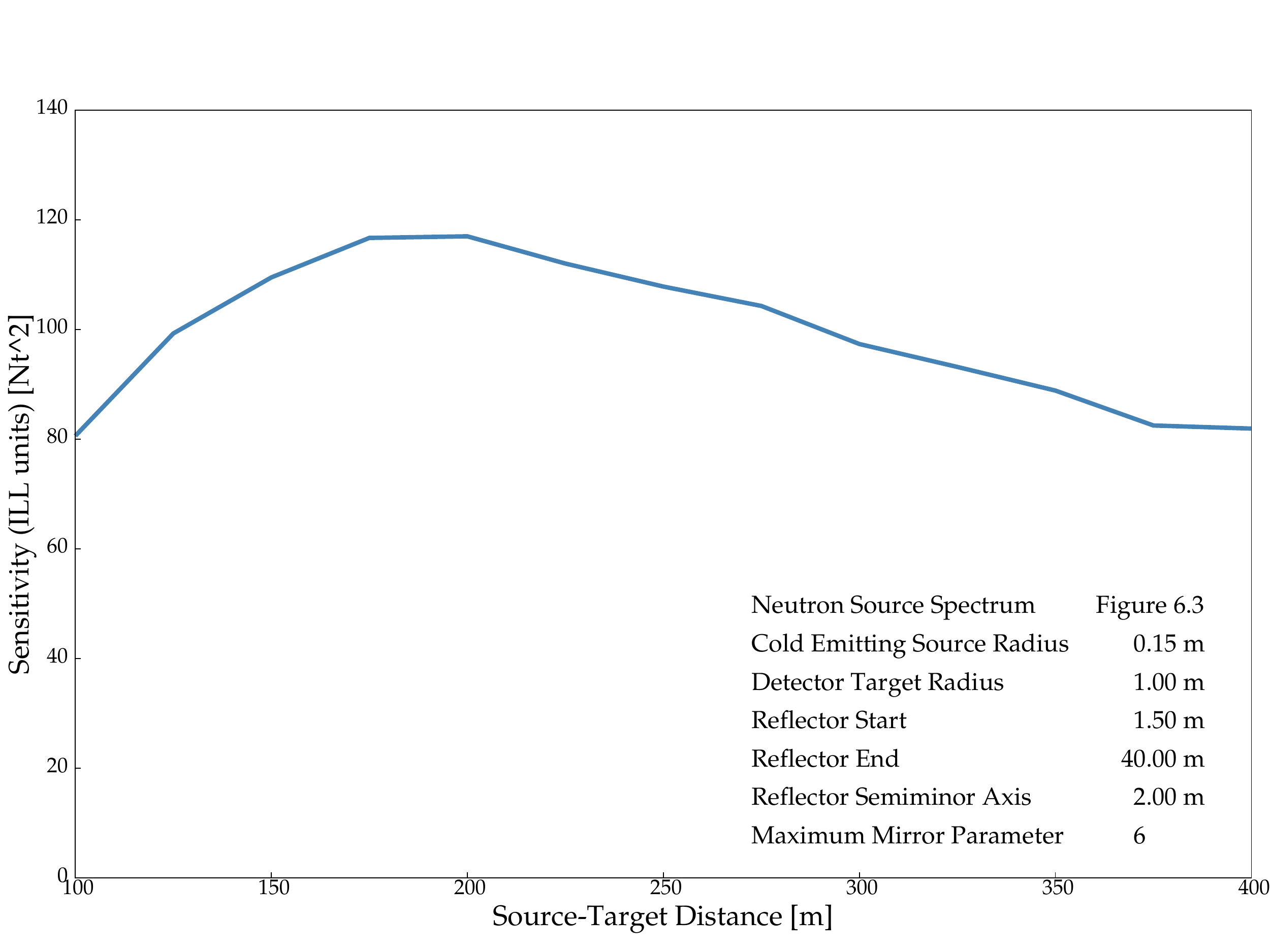}}
    \caption{a) Schematic diagram of a candidate NNbarX geometry, depicting the relative
location of the cold neutron source, reflector, target, and annihilation detector.  b)  Calculation of the $n$-$\bar{n}$ oscillation
sensitivity for a geometry similar to that in panel (a), where all parameters are fixed except for the source-target distance.}
    \label{nnbarx:fig:sensitivity}
\end{figure}

Supermirrors based on multi-layer coatings can greatly increase the
range of reflected transverse velocities relative to the nickel guides
used in the ILL experiment.  Supermirrors with $m = 4$, are now mass-produced and
supermirrors with up to $m = 7$, can be manufactured~\cite{Swiss:2013sn},
where ${\it m}$ is the ratio of the mirrorÕs critical angle for total external reflection to that of nickel (for a given wavelength).  To enhance the sensitivity of the $n$-$\bar{n}$ oscillation search, the
supermirrors can be arranged in the shape of a truncated focusing
ellipsoid~\cite{Kamyshkov:1995yk} (see Fig.~\ref{nnbarx:fig:sensitivity}a).
The focusing reflector with a large acceptance aperture will intercept neutrons
within a fixed solid angle and direct them by single reflection to the target.
The cold neutron source and annihilation target will be located in the focal planes of
the ellipsoid. The geometry of the reflector and the parameter ${\it
m}$ of the mirror material are chosen to maximize the sensitivity,
$N_{n}\cdot t^{2}$, for a given source brightness and a given
moderator and annihilation target size.  Elliptical
concentrators of somewhat smaller scale have already been implemented
for a variety of cold neutron
experiments~\cite{Boni:2010pb}.  The plan to develop a ${\it
dedicated}$ spallation neutron source for particle physics experiments
creates a unique opportunity to position the NNbarX neutron optical
system to accept a huge fraction of the neutron flux, resulting in
large gains in the number of neutrons directed to the annihilation
target.  Such a strategy makes use of a large fraction of
the available neutrons from the cold source, so it would be incompatible 
with a typical multi-user materials research facility, which would result in a reduction of $n$-$\bar{n}$ sensitivity.  Initial steps towards an optimized design have
been taken, with an NNbarX source design similar to the SINQ source
modeled and vetted vs. SINQ source performance (see
Fig. \ref{nnbarx:fig:source}), and a partially optimized elliptical
neutron optics system shown in Fig. \ref{nnbarx:fig:sensitivity}a.

A MCNPX~\cite{Mcnpx:1981mc} simulation of the performance of the cold
source shown in Fig.~\ref{nnbarx:fig:source} produced a flux of
cold neutrons emitted from the face of cryogenic liquid  deuterium
moderator into forward hemisphere with the spectrum shown in
Fig.~\ref{nnbarx:fig:source}.  Only a fraction of the integrated flux is
accepted by the focusing reflector to contribute to the sensitivity at
the annihilation target.  Neutrons emitted from
the surface of neutron moderator were traced through  the detector
configuration shown in Fig.~\ref{nnbarx:fig:source} with gravity
taken into account and with focusing reflector  parameters that were
adjusted by a partial optimization procedure. The flux of cold
neutrons impinging on the annihilation detector target located at the
distance $L$ from the source was calculated after reflection (mostly
single) from the focusing mirror. The time of flight to the target
from the last reflection was also recorded in the simulation
procedure. Each traced neutron contributed to the total
sensitivity figure $N_n\cdot t^{2}$ that was finally normalized to the
initial neutron flux from the moderator. Sensitivity as function of
distance between neutron source and target is shown in
Fig.~\ref{nnbarx:fig:sensitivity}(b). The simulation has several parameters
that affect the sensitivity: emission area of the moderator, distance
between moderator and annihilation target, diameter of the
annihilation target, starting and ending distance for truncated
focusing mirror reflector, semi-major axis of the ellipsoid ($L$/2), and the
reflecting value ``$m$" of the mirror.  Sensitivity is a complicated
functional in the space of these parameters. A vital element of
our ongoing design work is to understand the projected cost for the
experiment as a function of these parameters.

A sensitivity in NNbarX in units of the ILL experiment larger than 100
per year of running seems feasible from these simulations.  Configurations of
parameters that would correspond to even larger sensitivities are
achievable, but for the baseline simulation shown in
Fig.~\ref{nnbarx:fig:sensitivity} we have chosen a set of parameters that
we believe will be reasonably achievable and economical after
inclusion of more engineering details than can be accommodated in our
simulations to date.  The optimal neutron optical configuration for an
$n$-$\bar{n}$ oscillation search is significantly different from anything that has
been built, so the impact on the sensitivity of cost
and engineering considerations is not simple to
predict at such an early stage of the project. To demonstrate that the
key sensitivity parameters predicted by these
simulations do not dramatically depart from existing engineering
practice, we include Table~\ref{edm:tab:lqcd}, which shows the value of these same
parameters at existing MW-scale spallation neutron sources for the
source and optical parameters, and the ILL experiment for the
overall length $L$.

\begin{table}[ht]
\begin{threeparttable}
\centering
    	\caption{Comparison of parameters in NNbarX simulations with
    			existing practice.}
    	\label{edm:tab:lqcd}
    	\begin{tabular}{ccccc}
        \hline\hline Parameter & Units & Used in & Existing MW &
        References \\ & & Simulations & Facility Value & \\ \hline
        Source brightness & $n$/(s cm$^{2}$ sterad MW) &
        3.5$\times$10$^{12}$ & 4.5$\times$10$^{12}$ &
        ~\cite{Maekawa:2010fk} \\  ($E <$ 400 meV) &  &  &  & \\
        Moderator viewed area & cm$^{2}$ & 707 & 190 &
        ~\cite{Maekawa:2010fk} \\ Accepted solid
        angle\tnote{1} & sterad & 0.2 & 0.034 & ~\cite{Kai:2005tk} \\ Vacuum
        tube length & m & 200 & 100 & ~\cite{Baldo:1994bc} \\ $^{12}$C
        target diameter & m & 2.0 & 1.1 & ~\cite{Baldo:1994bc} \\
        \hline\hline
    \end{tabular}
    \begin{tablenotes}
    \item [1] Note that the solid angle quoted from JSNS
        is the total for a coupled parahydrogen moderator feeding 5
        neighboring beamlines (each of which would see a fifth of this
        value), whereas at NNbarX the one beam accepts the full solid
        angle.
    \end{tablenotes}
    \end{threeparttable}
\end{table}

\subsection{Requirements for an Annihilation Detector}
\label{nnbar:subsec:detector}

\begin{figure}[hbtp]
  \centering
    \subfloat[]{\includegraphics[width=3.8in]{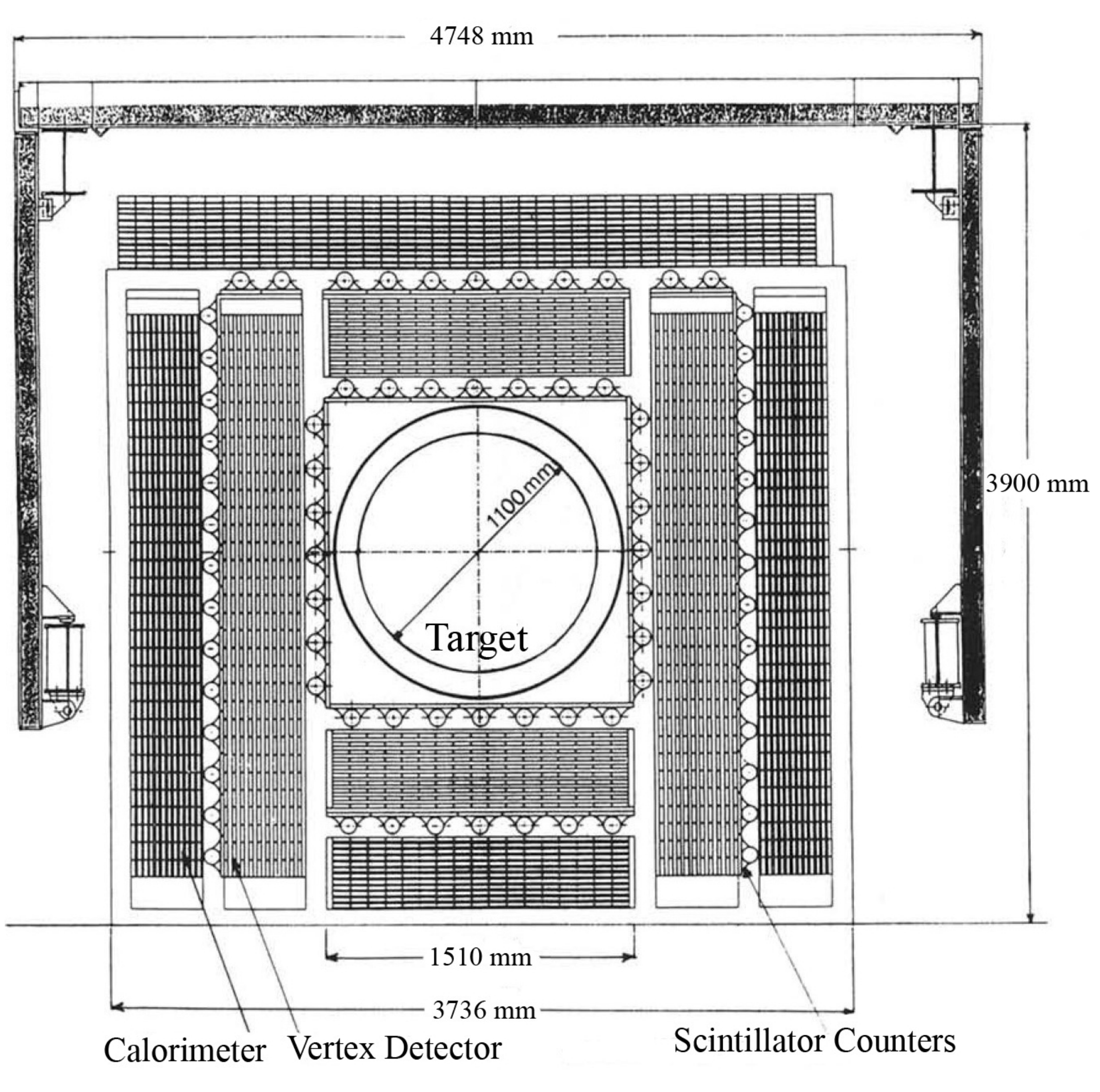}}
    \qquad
    \subfloat[]{\includegraphics[width=4.1in]{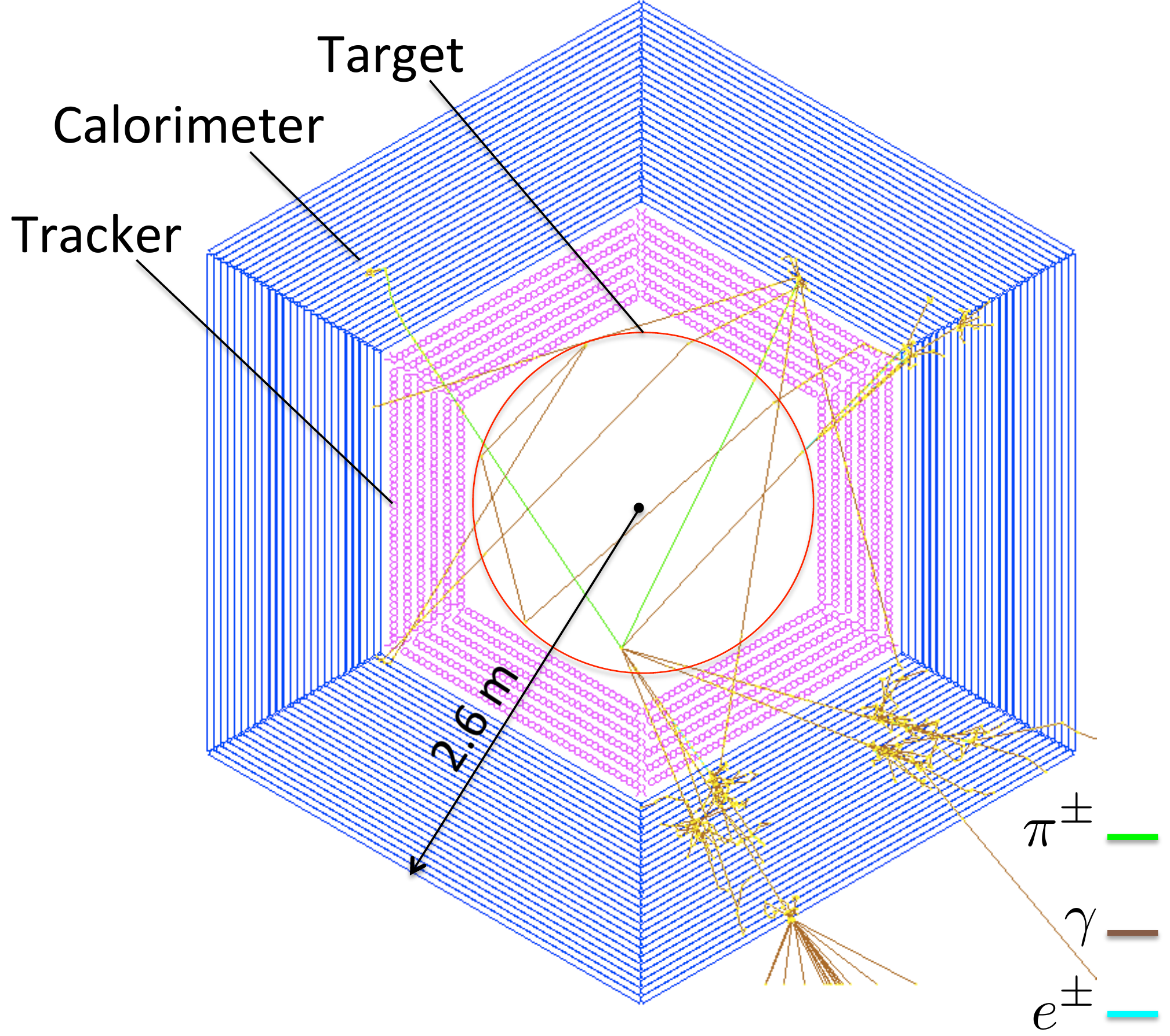}}
    \caption{a) Cross-sectional drawing of the ILL $n$-$\bar{n}$
    annihilation detector~\cite{Baldo:1994bc}.  b) Cross-sectional event display from our
    preliminary Geant4~\cite{Geant:2013ge} simulation for an annihilation event in a hexagonal NNbarX detector geometry with a length of 10 m.}
    \label{nnbarx:fig:detector}
\end{figure}

The target vacuum and magnetic field of 10$^{-5}$ Pa and $|{\vec B}| <$ 1 nT respectively is
achievable with standard vacuum technology and with an
incremental improvement on the ILL experiment through passive
shielding and straight-forward active field
compensation~\cite{Kronfeld:2013ak,Baldo:1994bc}.  In the design of the
annihilation detector, our strategy is to develop
a state-of-the-art realization of the detector design used in the ILL
experiment~\cite{Baldo:1994bc} (see Fig. \ref{nnbarx:fig:detector}a).  The
spallation target geometry of NNbarX introduces a new
consideration in the annihilation detector design, because of the
possible presence of fast neutron and proton backgrounds.  We defer discussion of the impact of the fast backgrounds on detector design to Sec.~\ref{nnbar:sec:backgrounds}, and concentrate here on our general detector design strategy.

In general, the $n$-$\bar{n}$ detector doesn't require premium performance, but
due to its relatively large size needs careful optimization of the
cost. In the current NNbarX baseline experiment, a uniform carbon disc in the
center of the detector vacuum region with a
thickness of $\sim$ 100 $\mu$m and diameter $\sim$ 2 m would serve as an annihilation
target.  Carbon is useful as an annihilation target due to the low capture cross
section for thermal neutrons $\sim$ 4 mb and high annihilation cross-section
$\sim$ 4 kb.  The fraction of hydrogen in the carbon film should be controlled
below $\sim$ 0.1$\%$ to reduce generation of capture $\gamma$'s.  The detector
should be built along a $\sim$ 4 m diameter vacuum region and cover a significant
solid angle in $\theta$-projection from $\sim$20$^{\circ}$ to 160$^{\circ}$ corresponding
to the solid angle coverage of $\sim$94$\%$.  The tube encompassing the neutron beamline vacuum  region should have a thickness
of $\sim$ 1.5 cm and be made of low-${\it Z}$ material (Al) to reduce multiple
scattering for tracking and provide a low (${\it n}$,$\gamma$) cross-section.
Additional lining of the inner surface of the vacuum region with $^{6}$LiF pads will
reduce the generation of $\gamma$'s by captured neutrons.  The detector vacuum
region is expected to be the source of $\sim$ 10$^{8}$ $\gamma$'s per second
originating from neutron capture. 

A tracker system should extend radially from the outer surface of the
detector vacuum tube by $\sim$ 50 cm.  It should provide
rms $\leq$ 1 cm accuracy for annihilation vertex reconstruction to the
position of the target in the $\theta$-projection (compared to 4 cm in
ILL experiment). This is a very important resource for the control of
background suppression in the detector.  Reconstruction accuracy in the
$\phi$-projection can be a factor of 3 - 4 lower than the $\theta$-projection and not degrade background rejection capability.  Relevant tracker
technologies can include straw tubes, proportional and drift
detectors.  A system similar to the ATLAS transition radiation tracker (TRT) is
currently under consideration for the tracking system.  Each straw tube in the ATLAS TRT
has an intrinsic coordinate resolution of 130 $\mu$m, which is more than adequate for the NNbarX annihilation detector.  We do, however, expect some optimization of these straw tubes will be required to adapt them to the NNbarX application.  For example, the
ATLAS TRT is capable of providing tracking for charged particles down
to a transverse momentum of $p_{T} =$ 0.25 GeV with an efficiency
of 93.6$\%$, but typically places a cut of $p_{T} >$ 1.00 GeV due to
combinatorics on the large number of tracks in collision events~\cite{Boldyrev:2012ab,Vankooten:2013rv}.  As with each candidate detector technology, an NNbarX optimization strategy will be identified and tested when possible. 

A time of flight (TOF) system should consist
of two layers of fast detectors (e.g. plastic scintillation slabs or tiles) before and after
the tracker.  With two layers separated by $\sim$50 cm - 60 cm, the TOF systems
should have timing accuracy sufficient to discriminate the annihilation-like tracks from the
cosmic ray background originating outside the detector volume.

The calorimeter will range out the annihilation products and should
provide trigger signal and energy measurements.  The average multiplicity of
pions in annihilation at rest equals 4.5, so an average pion can be
stopped in $\sim$20 cm of dense material (like lead or iron). For low
multiplicity (but small probability) annihilation modes, the amount of
material can be larger. The calorimeter configuration used in the ILL
experiment, with 12 layers of Al/Pb interspersed with gas detector
layers, might be a good approach for the
calorimeter design. Detailed performance for the measurement of total
energy of annihilation events and momentum balance in $\theta$- and
$\phi$-projections should be determined from simulations. An approach using
MINER$\nu$A-like wavelength shifting fibers coupled to extruded polystyrene scintillator is also being
considered~\cite{Mcfarland:2006km}.  

Although modern detector choices may eliminate the  need for a separate cosmic ray veto system (CVS) surrounding the calorimeter, we include it in our baseline design at present.  The cosmic veto system (CVS) surrounding the calorimeter should efficiently tag cosmic ray
background.  Large area detectors similar to MINOS scintillator
supermodules~\cite{Michael:2008dm} might be a good approach to the
configuration of the CVS. Possible use of timing information should be
studied in connection with the TOF system. 

\section{NNbarX Simulation}
\label{nnbar:sec:simulation}

Developing a detector model that allows us to reach our goal of zero background
and optimum signal event detection efficiency is the primary goal of our simulation campaign,
which is currently underway.  We are using Geant 4.9.6~\cite{Geant:2013ge} to simulate the
passage of annihilation event products through the annihilation detector geometry.  A detailed treatment of
$n$-$\bar{n}$ annihilation modes in $^{12}$C is currently under development.  According to
a Super-Kamiokande simulation study, 90$\%$ of the $n$-$\bar{n}$ annihilation modes in $^{16}$O
are purely pionic, while the remaining 10$\%$ are captured in the $\pi^{+}\pi^{-}\omega$
mode~\cite{Abe:2011ka}, which we expect to be similar to the physics of NNbarX.  The event
generator for $n$-$\bar{n}$ annihilation modes in $^{12}$C and fragmentation modes of the residual nucleus uses programs developed for the IMB experiment and Kamiokande II collaborations~\cite{Jones:1984tj,Takita:1986mt} validated in part by data from the Fr\'{e}jus, LEAR, and Super-Kamiokande experiments~\cite{Abe:2011ka,Golubeva:1996eg,Berger:1990cb,Fukuda:2003yf,Botvina:1990ab}.  The cross
sections for the $\pi$-residual nucleus interactions were based on extrapolation from measured
$\pi$-$^{12}$C and $\pi$-Al cross sections.  Excitation of the $\Delta$(1232) resonance was the
most important parameter in the nuclear propagation phase.  Nuclear interactions in the event
generator include $\pi$ and $\omega$ elastic scattering, $\pi$ charge exchange, $\pi$-production,
$\pi$-absorption, inelastic $\omega$-nucleon scattering to a $\pi$, and $\omega$ decays inside
the nucleus.  Fig. \ref{nnbarx:fig:detector}b shows an event display from our preliminary Geant4
simulation of an annihilation event in a detector geometry with a straw tube tracker and a calorimeter made of polystyrene scintillators and Pb absorbers.   

\section{NNbarX Backgrounds}
\label{nnbar:sec:backgrounds}

\begin{figure}[hbtp]
  \centering
    \subfloat[]{\includegraphics[width=5.0in]{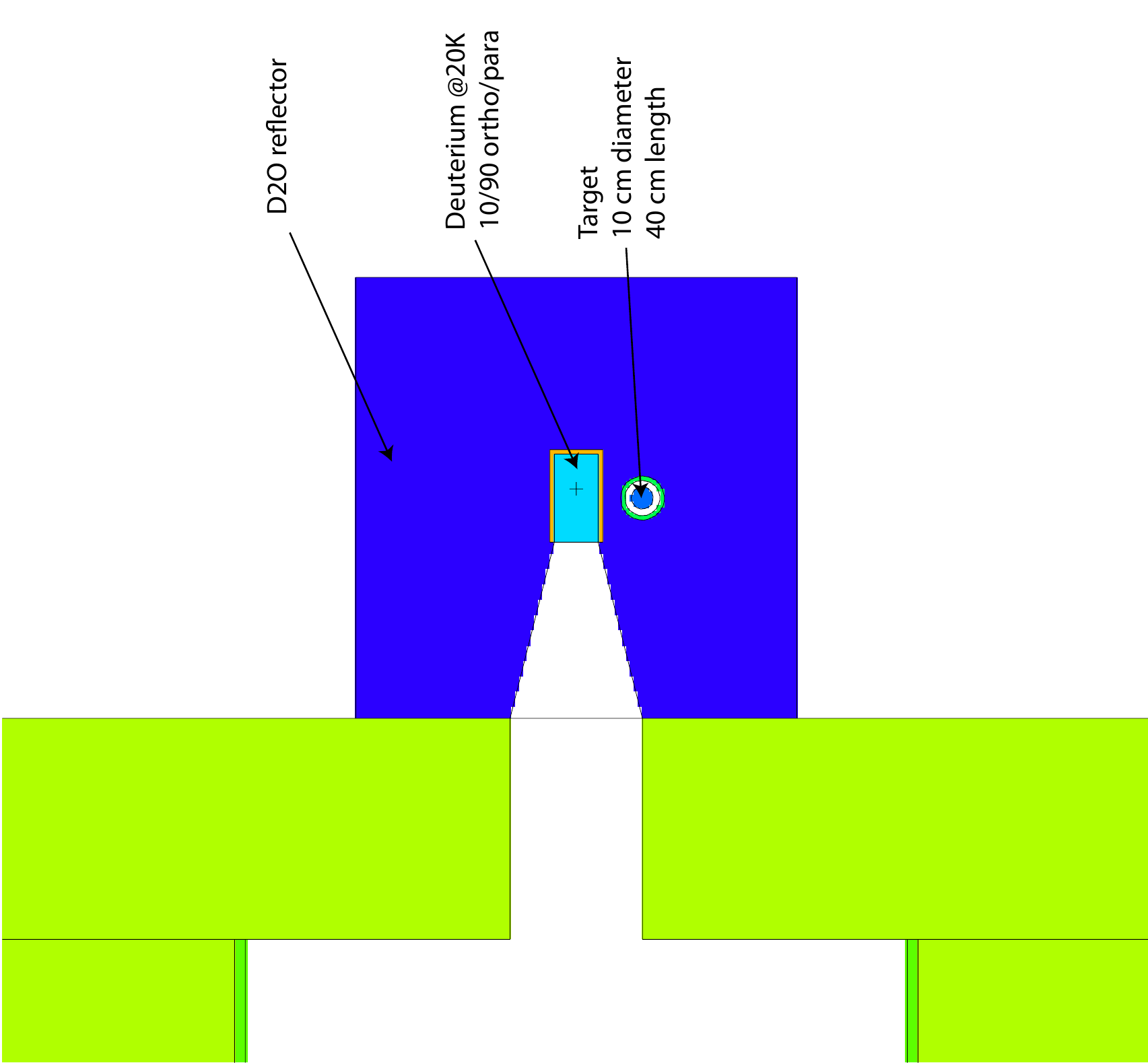}}
    \qquad
    \subfloat[]{\includegraphics[width=5.0in]{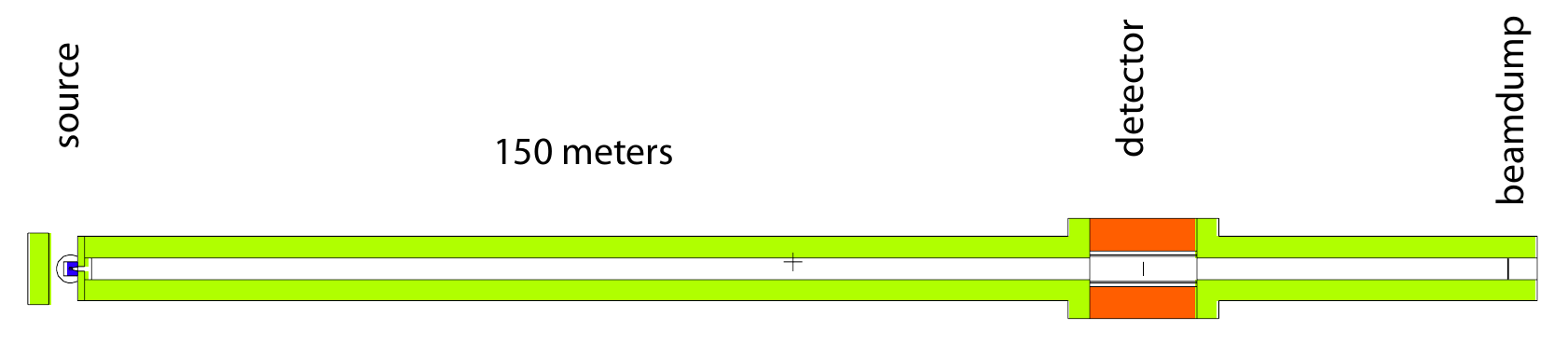}}
    \caption{a)  NNbarX source geometry in MCNPX.  The concrete walls along the beam line are 1 m thick.  b) Layout of the proposed NNbarX experiment in MCNPX with source, beam line, annihilation detector and beam dump.}
    \label{nnbarx:fig:mcnpxgeom}
\end{figure}

\begin{figure}[hbtp]
  \centering
    \subfloat[]{\includegraphics[width=4.5in]{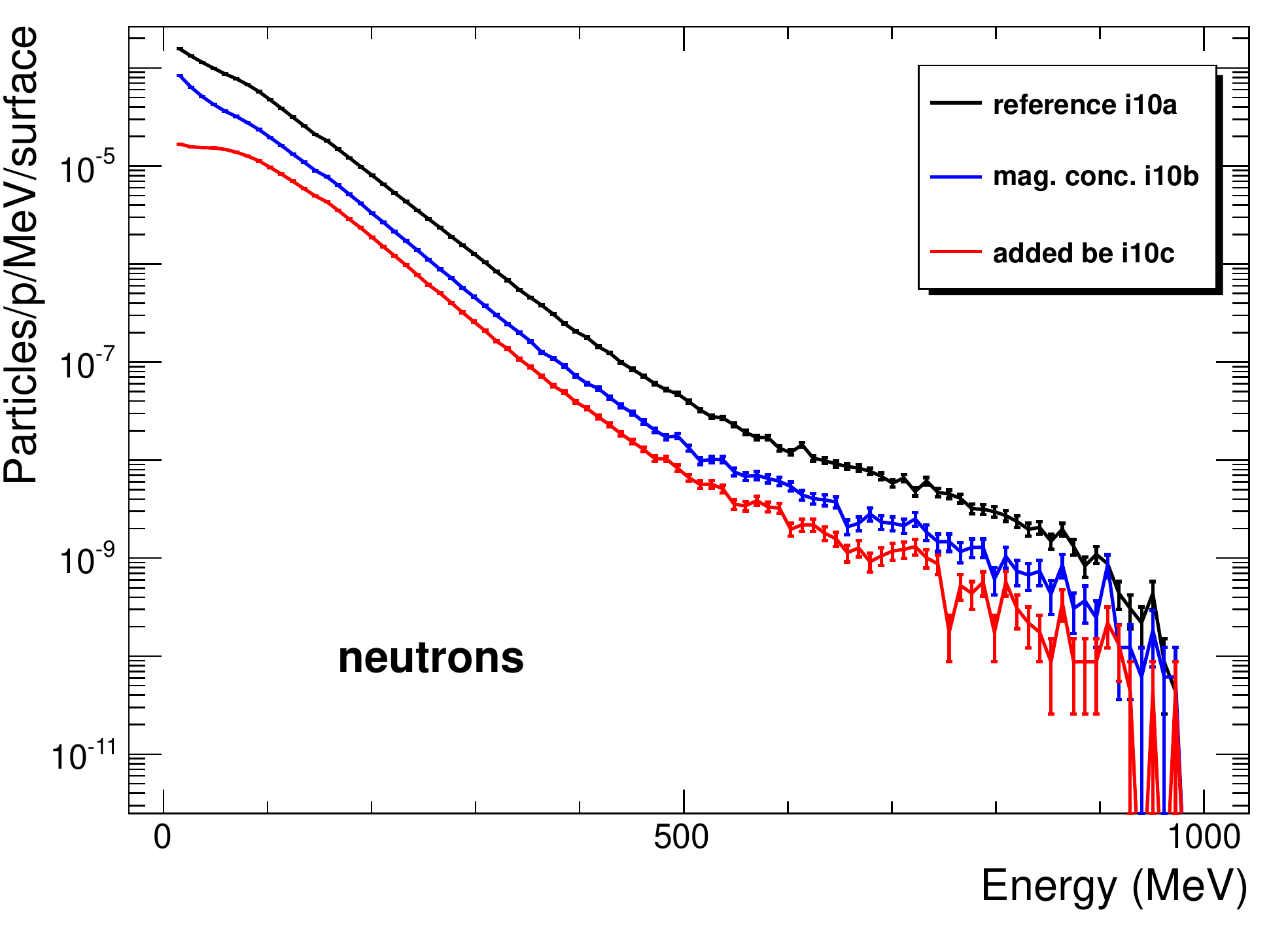}}
    \qquad
    \subfloat[]{\includegraphics[width=4.5in]{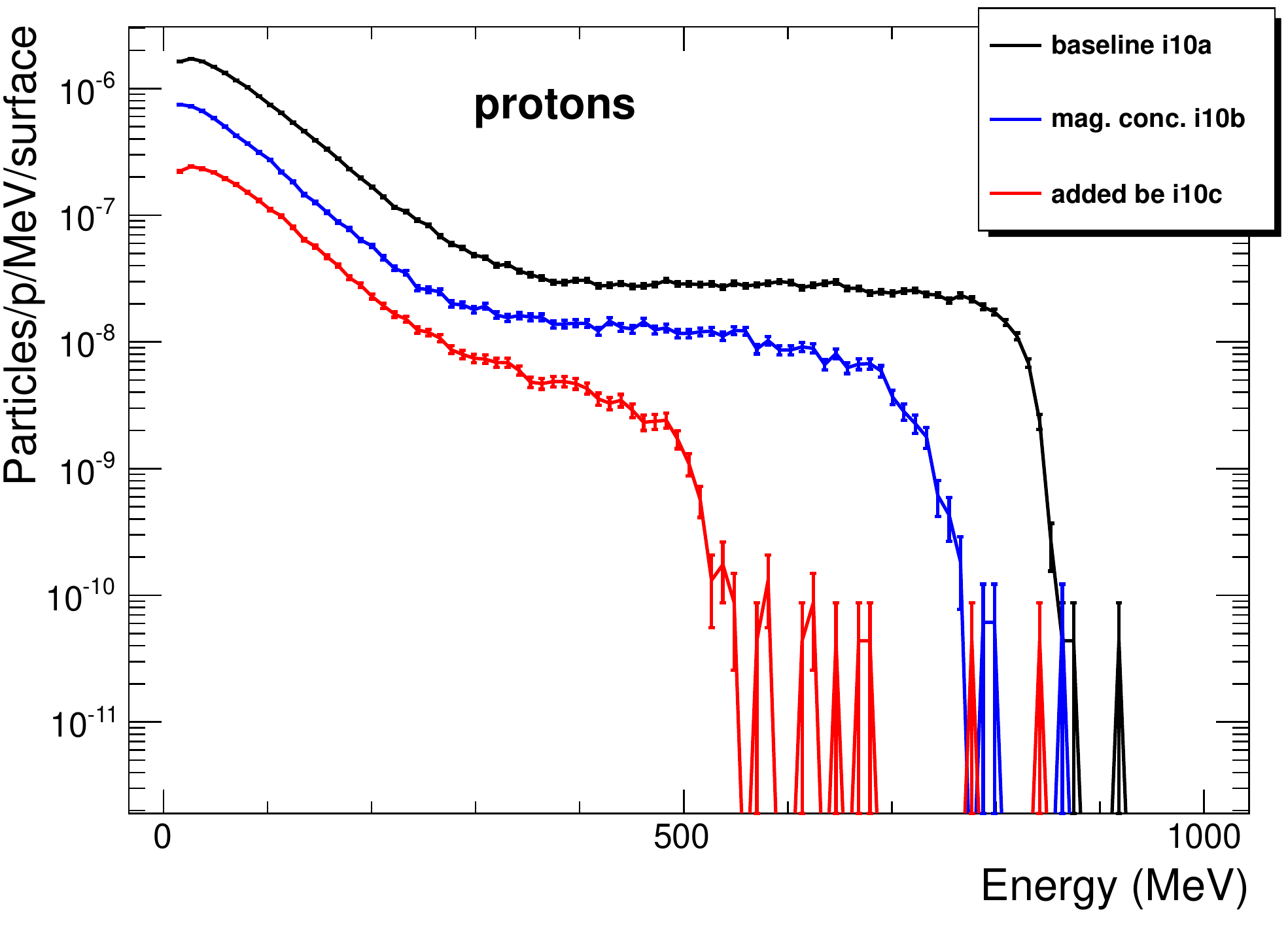}}
    \caption{MCNPX~\cite{Mcnpx:1981mc} simulation of the quasi-continuous production of a) neutrons and b) protons from a simplified spallation target geometry of Fig.~\ref{nnbarx:fig:mcnpxgeom}a (with a 1 GeV proton beam on target) after traversing a distance of 150 m through a totally evacuated volume (with no magnetic field) to the entrance of the annihilation detector region.  The black curve refers to the baseline configuration with concrete walls, while the blue curve is a configuration with magnetic concrete walls and the red curve is a simulation with a Be filter inserted in front of the spallation target.}
    \label{nnbarx:fig:fastbkgds}
\end{figure}

Although we base our overall approach to the annihilation event detector on the successful ILL detector configuration, the spallation target geometry of NNbarX introduces a new consideration in the annihilation detector design.  Initial simulations using a simplified spallation target geometry in MCNPX~\cite{Mcnpx:1981mc} (see Figs. \ref{nnbarx:fig:mcnpxgeom} and \ref{nnbarx:fig:fastbkgds}) and MARS~\cite{Mars:1998nm} indicate a possible presence of backgrounds from copious $\gamma$'s, fast neutrons and protons scattered from the spallation target.  Simulations using the spallation target geometry described in Sec.~\ref{nnbar:subsec:overview} are currently underway.  

The $\gamma$ backgrounds can be effectively cut with a veto on the detector correlated with the beam timing.  According to simulations in MCNPX~\cite{Mcnpx:1981mc} using the simplified spallation target geometry, a Be filter inserted in front of the spallation target will eliminate fast protons above 550 MeV (see Fig. \ref{nnbarx:fig:fastbkgds}b). The arrival time distribution at the detector for fast neutrons and the remaining fast protons will have a much larger spread, and may impact the design of the detector.  For NNbarX, we utilize a strategy of integrating our shielding scheme for
fast particles into the design of the source and beamline, and
optimize the choice of tracker detectors to differentiate between
charged and neutral tracks.  These backgrounds can be effectively eliminated with a slow proton beam-chopping protocol (1 ms on, 1 ms off), at the expense of roughly a factor of two in integrated neutron flux, making an improved background rejection strategy highly desirable.

\begin{figure}[hbtp]
  \centering
    \subfloat[]{\includegraphics[width=3.9in]{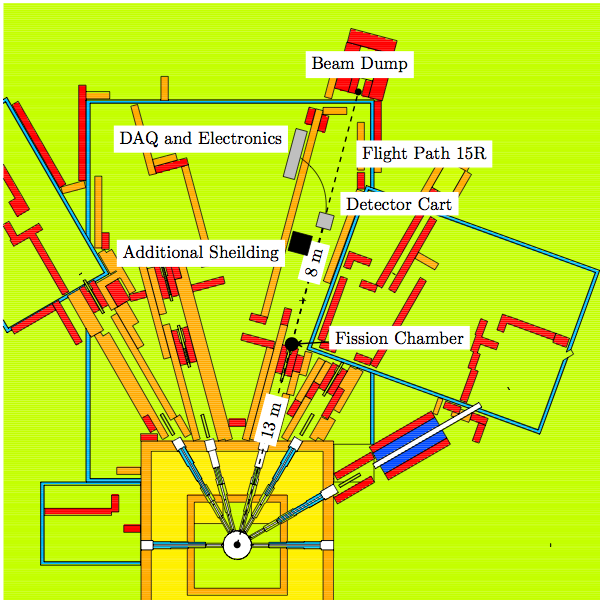}}
    \qquad
    \subfloat[]{\includegraphics[width=4.9in]{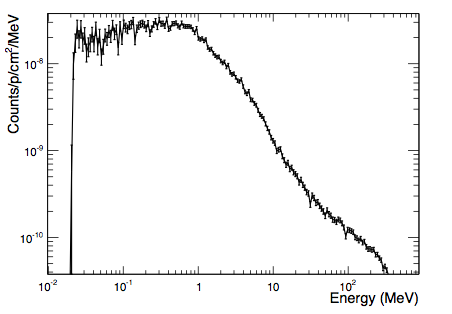}}
    \caption{a) LANSCE WNR-15R beam line. b) MCNPX-predicted neutron flux 20 m from the WNR target.}
    \label{nnbarx:fig:lanscewnr}
\end{figure} 

Currently, the response of tracker and scintillator technologies in experiments at the Intensity Frontier to fast neutrons is not well characterized.  Understanding the response of the NNBarX candidate tracker and calorimeter technologies to fast neutrons would be an important tool for background reduction in NNBarX and extremely useful for detector development in the search for ${\it n} -{\it \overline{n}}$ oscillation using free neutrons.  In order to provide a detailed understanding of our candidate detector technologies to fast neutrons, we have begun a characterization campaign at the WNR facility at Los Alamos National Laboratory (LANL).  The 800 MeV proton linear accelerator at the Los Alamos Neutron Science Center (LANSCE) provides a pulsed ($\approx$4 $\mu$A at 40 Hz) beam to the spallation target at the WNR facility (Fig. \ref{nnbarx:fig:lanscewnr}a).  Beamlines at fixed angles with respect to the incident proton beam are equipped with shutters and collimators, to provide a maximum fast neutron beam intensity of about 10$^{6}$ n/sec in a 5 cm diameter beam.  A key feature of the WNR facility is that a calibrated set of fission-foil detectors continuously monitors each fast neutron beam~\cite{Wender:1993sw}, providing the absolute neutron flux as a function of energy.  According to a 2012 test run, the pulse beam structure from the LANSCE accelerator allows for the neutron energy to be determined via the time of flight, with an energy resolution of $\simeq$ 0.4$\%$ at a few MeV to $\simeq$ 8$\%$ at 800 MeV.  According to simulations in MCNPX~\cite{Mcnpx:1981mc}, the LANSCE WNR beam line will provide a neutron energy spectrum similar to the expected neutron energy distribution in Project X (Fig. \ref{nnbarx:fig:lanscewnr}b).  Ultimately, the measured fast neutron response for candidate technologies will be integrated into our detector simulation campaign to permit a realistic assessment of their impact to our sensitivity and background rejection procedure.

\section{The NNbarX Research and Development Program}
\label{nnbar:sec:randd}

In October of 2012, the FNAL Physics Advisory Committee strongly
supported the physics of NNbarX and recommended that ``R$\&$D be
supported, when possible, for the design of the spallation target, and
for the overall optimization of the experiment, to bring it to the
level required for a proposal to be prepared.''   The NNbarX collaboration
has identified several areas where research and development may
substantially improve the physics reach of the experiment and/or significantly reduce its costs: target and
moderator design, neutron optics optimization and the annihilation
detector design.  At the core of this
activity is an integrated models for the source, neutron optics and
detectors that provides a useful tool for evaluating overall sensitivity to
annihilation events and backgrounds (particularly from fast neutrons), and for developing a cost
scaling model.  

There exist a number of improvements for the target and moderator as discussed in Sec.~\ref{nnbar:subsec:sensitivity}, which have already
been established as effective and might be applied to our baseline
conventional source geometry.  For example, one can shift from a ${\it
cannelloni}$ target to a lead-bismuth eutectic (LBE)
target~\cite{Wagner:2008ww}, utilize a re-entrant moderator
design~\cite{Ageron:1989pa}, and possibly use
reflector/filters~\cite{Mocko:2013mm}, supermirror
reflectors~\cite{Swiss:2013sn}, and high-albedo materials such as
diamond nanoparticle
composites~\cite{Nezvizhevsky:2008vz,Lychagin:2009el,Lychagin:2009em}.
At present, the collaboration envisions a program to perform neutronic
simulations and possibly benchmark measurements on several of these
possibilities, with high-albedo reflectors as a priority.  A collaboration that includes researchers from Indiana University (IU), the Spallation Neutron Source, LANL, and the European Spallation Source expects to be investigating these and other similar neutronic questions over the next several years using the LENS facility at IU~\cite{Lavelle:2008cl}.  Although
the basic performance of neutron optics is established, optimizing the selection of
supermirror technology for durability (versus radiation damage) and cost
could have a very large impact on the ultimate reach of the experiment.  We note that advances in source efficiency and the performance of neutron optics could also be used to substantially reduce the cost of launching this experiment by reducing the accelerator performance required to reach our design sensitivity goals. These same advances will, of course, also be important to any future fundamental physics experiment relying on intense beams of neutrons.  Our international collaborators from Japan and India are involved in the development of technology for cost reduction of high-m super-mirrors, in the economical design of active and passive magnetic shielding, and in the study and prototyping of possible detector options.  We plan to explore the possibility of employing existing neutron production facilities in the country that could allow for a reduction in the cost of a $n$-$\bar{n}$ oscillation search experiment.

As discussed in Sec.~\ref{nnbar:sec:backgrounds}, the collaboration is currently using the WNR facility
at LANSCE to determine the detection efficiency and timing properties
of a variety of detectors from a few MeV to 800 MeV neutrons.  Detectors
under evaluation include carbon fiber-body proportional gas tubes, straw tubes, and extruded
polystyrene scintillators.  Characterizing different
available detector options and modernizing the annihilation detector
should improve the background rejection capability and permit reliable
scaling to more stringent limits for $n$-$\bar{n}$ oscillations.  Given that fast neutron backgrounds are extremely difficult to shield, can produce very energetic events in detectors, and have a reasonable likelihood of only being partially contained, the sensitivity of various detector types to fast neutrons is an issue in many proposed experiments of relevance to the intensity frontier.  Therefore, we see this activity as having an impact far beyond the scope of $n$-$\bar{n}$ oscillations.  Although many promising avenues to improve sensitivity have been identified, it is also recognized that one of the main technical challenges for NNbarX is to minimize the cost of
critical hardware elements, such as the large-area super-mirrors,
large-volume magnetic shielding, vacuum tube, shielding of the
high-acceptance front-end of the neutron transport tube, and
annihilation detector components.  Reconciling the need to minimize cost and optimize sensitivity is the central goal of the current design activity for the NNbarX experiment.

Finally, as discussed in Sec.~\ref{nnbar:sec:simulation}, we will continue development of software tools for neutron source and annihilation target simulation and perform a multi-dimensional optimization of parameters.  This optimization includes the cost as well as the refinement of the event generators relevant to antineutron annihilation within nuclei for both free and intranuclear $n$-$\bar{n}$ searches (including $^{16}$O, $^{40}$Ar) . 
\section{Summary}
\label{nnbar:sec:summary}

Assuming that beam powers up to 1 MW on the spallation target and that 1
GeV protons are delivered from the Project X linac, the goal of NNbarX
will be to improve the sensitivity of an $n$-$\bar{n}$ search
($N_{n}\cdot t^{2}$) by at least a factor 30 per year of running (compared to the previous
limit set in ILL-based experiment~\cite{Baldo:1994bc}) with a
horizontal beam experiment; and by an additional factor of $\sim$ 100
in a second stage with the vertical layout. The R$\&$D phase of the
experiment, including development of the conceptual design of the cold
neutron spallation target, and conceptual design and optimization of
the performance of the first-stage of NNbarX is expected to take 2-3
years.  Preliminary results from this effort suggest that an
improvement over the ILL experiment by a factor of more than 100 in sensitivity may
be realized even in this horizontal mode, but more work is needed to
estimate the cost of improvements at this level.  The running time of
the first stage of the NNbarX experiment is anticipated to be 3 years. The
second stage of NNbarX will be developed depending upon the
demonstration of the technological principles and techniques of the first
stage.


\bibliographystyle{apsrev4-1} \bibliography{Snowmass_nnbar}
\end{document}